\documentclass[aps,prl,twocolumn,groupedaddress]{revtex4}
\usepackage{amsmath}
\usepackage{amsfonts}

\def\bF{\mathbf F}
\def\bE{\mathbf E}
\def\bH{\mathbf H}
\def\bJ{\mathbf J}
\def\bB{\mathbf B}
\def\bD{\mathbf D}

\def\bX{\mathbf X}
\def\bY{\mathbf Y}

\def\bx{\mathbf x}

\def\bk{\mathbf k}

\def\b0{\mathbf 0}

\begin{document}

\title{Collapsing Solutions of the Maxwell Equations}
\author{Neil V. Budko}
\affiliation{Laboratory of Electromagnetic Research, Faculty of Electrical Engineering, Mathematics and Computer Science,
Delft University of Technology,
Mekelweg 4, 2628 CD Delft, The Netherlands}
\email{n.budko@ewi.tudelft.nl}
\author{Alexander B. Samokhin}
\affiliation{Department of Applied Mathematics, Moscow Institute of
Radio Engineering, Electronics, and Automatics (MIREA), Verndasky~av.~78, 117454, Moscow, Russian Federation}
\thanks{This research is supported by NWO (The Netherlands)}

\date{\today}

\begin{abstract}
We derive the essential space-time spectrum of the Maxwell equations in linear isotropic inhomogeneous media 
together with the corresponding essential modes. These modes represent the collapse of the electromagnetic 
field into a single point in space at a single angular frequency. The location and frequency of the essential mode 
are random variables obeying the Born statistical postulate.
\end{abstract}


\maketitle
Wavefunction collapse is one of the phenomenological postulates of the orthodox quantum theory. 
However, this concept is so foreign to the wave behavior in general, that many find it impossible to accept.
The recent years have seen an increase in the number of publications devoted to the foundations of quantum 
physics. The more constructive of the new interpretations suggest that the collapse is just an awkward 
theoretical construct, claiming that there is no such thing in reality. 

We mention two such theories that seem to have passed the test of time and are often regarded as candidates for a new, 
less orthodox, model (see \cite{Schlosshauer2005} for an overview). One, called the decoherence theory, operates almost entirely 
within the standard postulates of quantum mechanics, 
including the statistical interpretation of the wavefunction. The decoherence theory argues that 
the most important wave feature of the quantum behavior is the interference and as such it can only be observed with coherent wavefunctions. 
Therefore, the apparent breakdown of wave behavior in the process of measurement and, 
more generally, in macroscopic objects, is nothing else but the major loss of coherence induced by the environment. 
The loss of coherence, however, is never perfect, so that some interference is always present, even at a macroscopic level.
Isolated microscopic objects would then never experience collapse at all. On the other hand, may be there is no such thing
as a truly isolated microscopic object.

Another theory, known as the dynamical reduction model, suggests that the basic wave equation of quantum 
mechanics -- the Schr{\"o}dinger equation -- should be modified, so that it would admit collapsing solutions. 
The proposed modifications include a nonlinear and/or a stochastic term, since it is believed that a unitary linear wave evolution 
can never exhibit a random collapsing behavior. The dynamical reduction program neither provides nor aims at 
a perfectly collapsing wavefunction, and imposes the quantum statistics a priori.
Hence, both the decoherence theory and the dynamical reduction model introduce something that ``looks and smells like a collapse''
\cite{TegmarkWheeler2001}, but in our view is not the real thing. 

We believe that the true problem with accepting the orthodox point of view is largely due to the absence of adequate 
mathematics that would describe collapse of the wave behavior at least in some particular setting. 
We shall not go into the details of the two interpretations mentioned above, 
which question the very existence of the collapse. Instead, we shall provide an example 
of a perfect collapse of wave motion, which has nothing to do with coherence or nonlinearity. 
Namely, we shall show that the linear Maxwell's equations admit generalized collapsing solutions, 
that are not only random, but also satisfy the Born statistical postulate in a certain sense.

A few words have to be said about the mathematical techniques employed here. We follow up on the discovery of the
essential spectrum in the volume integral formulation of the electromagnetic scattering problem \cite{BudkoSamokhin2006a}
derived using the symbol theory of Mikhlin and Pr{\"o}ssdorf \cite{Mikhlin2}. 
Our preliminary qualitative analysis indicated that the ``essential resonance'', which happens if the essential spectrum 
gets close to zero, must be related to the so-called plasmon, where the electromagnetic field is coupled 
to the plasma wave and is highly localized at a plasma-dielectric interface \cite{BudkoSamokhin2006b}.
This prompted further investigation and recently we were able to prove that the mode corresponding to the point of essential spectrum, called singular mode, is
a very special function, indeed \cite{BudkoSamokhin2006c}. We have used the Weyl definition of spectrum \cite{HislopSigal1996} 
showing that the singular mode is a distribution, which is best described as the square root of the 
Dirac delta function. Such functions obey the Colombeau algebra \cite{Colombeau1992} and, despite their unique
properties, are not in wide use today.

Here we apply the Weyl definition of spectrum to the differential Maxwell's equation and show that the electromagnetic essential
spectrum is not a unique feature of the volume integral formulation only. In fact, the spatial singular modes, 
derived for the integral operator in \cite{BudkoSamokhin2006c}, will appear in the differential case as well. To arrive at a 
physically interpretable picture we do not consider the spatial and temporal spectra separately,
but immediately derive the essential space-time spectrum of the Maxwell's operator. 
Based on the unique structure of the essential modes, their natural randomness, and Born-type statistical properties, 
we conclude that we deal here with the perfect collapse of the electromagnetic wave motion. 

Consider the Maxwell equations
 \begin{align}
 \label{eq:Maxwell}
 \begin{split}
 -\nabla\times \bH +\partial_{t}\bD&=-\bJ,
 \\
 \nabla\times \bE +\partial_{t}\bB&=\b0,
 \end{split}
 \end{align}
with linear isotropic constitutive relations
$\bD(\bx,t)=\varepsilon(\bx,t)*\bE(\bx,t)$,
$\bB(\bx,t)=\mu(\bx,t)*\bH(\bx,t)$,
where $\varepsilon(\bx,t)$ and $\mu(\bx,t)$ are some continuous functions with finite 
spatial support, and star denotes the temporal convolution. In the matrix form these equations can be written as
 \begin{align}
 \label{eq:MaxwellMatrix}
 \left[
 \begin{array}{cc}
 \partial_{t}\varepsilon* & -\nabla\times \\ 
 \nabla\times & \partial_{t}\mu*
 \end{array}
 \right]
 \left[
 \begin{array}{c}
 \bE\\
 \bH
 \end{array}
 \right]
 =
 \left[
 \begin{array}{c}
 -\bJ\\
 \b0
 \end{array}
 \right].
 \end{align}
In operator notation we simply write
 \begin{align}
 \label{eq:MaxwellOperator}
 {\mathbb M}\bX = \bY,
 \end{align}
where ${\mathbb M}$ is the Maxwell operator, and $\bX$ and $\bY$ are six-vectors.
We introduce the Hilbert space with the following norm:
 \begin{align}
 \label{eq:Norm}
 \left\Vert \bX \right\Vert^{2} = \int_{-\infty}^{\infty}\int_{\bx\in{\mathbb R}^{3}}
 \left\vert\bX\right\vert^{2}
 \,{\rm d}\bx\,{\rm d}t.
 \end{align}
According to the Weyl definition of spectrum of 
${\mathbb M}$ one needs to find the sequence of functions $\{\bF_{n}\}$ with the following properties:
 \begin{align}
 \label{eq:Normalization}
 \left\Vert\bF_{n}\right\Vert&=1,
 \\
 \label{eq:Minimization}
 \lim\limits_{n\rightarrow\infty}\left\Vert{\mathbb M}\bF_{n}-\lambda\bF_{n}\right\Vert&=0
 \end{align}
If the Weyl sequence that satisfies these equations for some $\lambda$ has no convergent subsequence,
then such $\lambda$ is in the essential spectrum. If for some other $\lambda$ the Weyl sequence does converge to 
a function from the Hilbert space, then this function is an eigenfunction, and $\lambda$ is an eigenvalue of ${\mathbb M}$.

The singular modes of the volume integral operator found in \cite{BudkoSamokhin2006c} are the following 
vector-valued functions:
\begin{align}
\label{eq:DefPsi}
\begin{split}
&\Psi(\alpha,\bx,\bx_{\rm c})=
\\
&\left(\frac{2}{3}\right)^{1/2}\pi^{-3/4}\alpha^{5/4}(\bx-\bx_{\rm c}) 
\exp\left(-\frac{\alpha}{2}\vert\bx-\bx_{\rm c}\vert^{2}\right),
\end{split}
\end{align}
where $\bx,\bx_{\rm c}\in{\mathbb R}^{3}$ and $\alpha\ge 0$ is the sequence parameter. 
The properties of these functions are summarized in Theorem~2.1 of \cite{BudkoSamokhin2006c}. 
The most important are the following two:
 \begin{align}
 \label{eq:NormalizedPsi}
 \int_{\bx\in{\mathbb R}^{3}}\left\vert\Psi(\alpha,\bx,\bx_{\rm c})\right\vert^{2}\,{\rm d}\bx&=1,
 \\
 \label{eq:Sifting}
 \lim\limits_{\alpha\rightarrow\infty}\int_{\bx\in{\mathbb R}^{3}}
 f(\bx)\left\vert\Psi(\alpha,\bx,\bx_{\rm c})\right\vert^{2}\,{\rm d}\bx &= f(\bx_{\rm c}).
 \end{align}
This is why we refer to singular modes as the square root of the delta function. Further we note that
the vector-valued $\Psi$ has the form of a position vector $(\bx-\bx_{\rm c})$ multiplied by a scalar, 
and its Fourier transform has the form of an angular vector $\bk$ multiplied by a scalar \cite{BudkoSamokhin2006c}. 
This means that the action of a curl operator on $\Psi$ gives zero for any value of parameters $\alpha$ and $\bx_{\rm c}$, 
which is the main reason why singular modes are
possible with the first-order Maxwell's equation, but not with the second-order wave equations in homogeneous media, 
where the spatial operator is the Laplacian.

Here we intend to take into account the temporal essential spectrum as well. Consider 
a similar sequence of scalar functions:
 \begin{align}
 \label{eq:DefPhi}
 \begin{split}
 &\phi(\beta,t,\omega_{\rm c})=
 \\
 &i\sqrt{2}\pi^{-1/4}\beta^{-3/4}t
 \exp\left(-\frac{1}{2\beta}t^{2}-i\omega_{\rm c}t\right).
 \end{split}
 \end{align}
The Fourier transform of these functions is
 \begin{align}
 \label{eq:DefPhiOmega}
 \begin{split}
 &\hat{\phi}(\beta,\omega,\omega_{\rm c})=
 \\
 &\sqrt{2}\pi^{-1/4}\beta^{3/4}(\omega-\omega_{\rm c})
 \exp\left(-\frac{\beta}{2}\vert\omega-\omega_{\rm c}\vert^{2}\right),
 \end{split}
 \end{align}
and, as can be verified using the integration techniques developed in \cite{BudkoSamokhin2006c}, it is normalized and 
also represents the square root of the Dirac delta function, but now in $\omega$-domain, i.e.,
 \begin{align}
 \label{eq:NormalizedPhi}
 \int_{-\infty}^{\infty}\left\vert\hat{\phi}(\beta,\omega,\omega_{\rm c})\right\vert^{2}\,{\rm d}\omega&=1,
 \\
 \label{eq:SiftingPhi}
 \lim\limits_{\beta\rightarrow\infty}\int_{-\infty}^{\infty}
 g(\omega)\left\vert\hat{\phi}(\beta,\omega,\omega_{\rm c})\right\vert^{2}\,{\rm d}\omega &= g(\omega_{\rm c}).
 \end{align}
The product of functions (\ref{eq:DefPsi}) and (\ref{eq:DefPhi}) can be used to generate the essential 
modes of the complete Maxwell's operator, i.e.,
 \begin{align}
 \label{eq:EssentialModes}
 \bF^{\rm e}=
 \left[
 \begin{array}{c}
 \phi\Psi\\
 \b0
 \end{array}
 \right],
 \;\;\;\;\;\;
 \bF^{\rm m}=
 \left[
 \begin{array}{c}
 \b0\\
 \phi\Psi
 \end{array}
 \right]. 
 \end{align}
Substituting $\bF^{\rm e}$ and $\bF^{\rm m}$ in (\ref{eq:Minimization}) we arrive at
 \begin{align}
 \label{eq:CheckFe}
 \lim\limits_{\alpha,\beta\rightarrow\infty}
 \left\Vert{\mathbb M}\bF^{\rm e}-\lambda^{\rm e}\bF^{\rm e}\right\Vert^{2}&=
 \left\vert\lambda^{\rm e}+i\omega_{\rm c}\hat{\varepsilon}(\bx_{\rm c},\omega_{\rm c})\right\vert^{2},
 \\
 \label{eq:CheckFm}
 \lim\limits_{\alpha,\beta\rightarrow\infty}
 \left\Vert{\mathbb M}\bF^{\rm m}-\lambda^{\rm m}\bF^{\rm m}\right\Vert^{2}&=
 \left\vert\lambda^{\rm m}+i\omega_{\rm c}\hat{\mu}(\bx_{\rm c},\omega_{\rm c})\right\vert^{2},
 \end{align}
meaning that $-i\omega_{\rm c}\hat{\varepsilon}(\bx_{\rm c},\omega_{\rm c})$ and 
$-i\omega_{\rm c}\hat{\mu}(\bx_{\rm c},\omega_{\rm c})$, where $\omega_{\rm c}\in{\mathbb R}$ and 
$\bx_{\rm c}\in{\mathbb R}^{3}$, are in the essential spectrum of the Maxwell operator.

Obviously, the Weyl definition of spectrum allows working with a very broad class of functions,
and it may seem that the modes of essential spectrum do not have any mathematical 
and as a result any physical meaning. Indeed, we show in \cite{BudkoSamokhin2006c}, that the singular modes 
do not (strongly) converge to any function of the Hilbert space. On the other hand, the square of these functions generates 
the generalized Dirac's delta function, which in its turn acquires some meaning only upon integration. All this amounts to saying 
that the essential modes may allow for some physical interpretation only if we square and integrate them. 

The physical quantity 
obtained by squaring and integrating the electromagnetic field over space and time has the 
dimensions of action [J\,s], the electric-field part of which is:
 \begin{align}
 \label{eq:Action}
 \begin{split}
 S^{\rm e}&=\int_{-\infty}^{\infty}\int_{\bx\in{\mathbb R}^{3}}
 \bE(\bx,t)\cdot\bD(\bx,t)
 \,{\rm d}\bx\,{\rm d}t
 \\
 &=\int_{-\infty}^{\infty}\int_{\bx\in{\mathbb R}^{3}}
 {\rm Re}\left[\hat{\varepsilon}(\bx,\omega)\right]\left\vert\hat{\bE}(\bx,\omega)\right\vert^{2}
 \,{\rm d}\bx\,{\rm d}\omega.
 \end{split}
 \end{align}
Here we have used the power (Plancherel) theorem of Fourier analysis, the fact that $\bE(\bx,t)$ is a real 
function of time, and the causality-driven assumption that the real part of $\hat{\varepsilon}(\bx,\omega)$ is
an even function of $\omega$, while the imaginary part is odd. Of course, $S^{\rm e}$ is a slightly artificial quantity.
It only serves to illustrate our point and techniques.

It is clear that the same numerical value of action may correspond to different states of the electric field.
For example, field $\hat{\bE}(\bx,\omega)$ may be a ``proper'' solution of the Maxwell equations.
Alternatively, it can be an essential mode, which satisfies Maxwell's equation only in the Weyl sense, but
produces a perfectly well-defined action. Consider a hypothetical transition between such states. 

Let the action $S^{\rm e}_{1}$
in the first state be represented by an integral over the electric field $\hat{\bE}_{1}$ satisfying the Maxwell 
equation in the usual sense, i.e. $S^{\rm e}_{1}$ is given by (\ref{eq:Action}) with $\hat{\bE}_{1}$ instead of $\hat{\bE}$.
In the second state the action is represented by a single essential mode
 \begin{align}
 \label{eq:EssentialAction}
 \begin{split}
 S^{\rm e}_{2}&\,= C_{0} \lim\limits_{\alpha,\beta\rightarrow\infty}\int_{-\infty}^{\infty}\int_{\bx\in{\mathbb R}^{3}}
 {\rm Re}\left[\hat{\varepsilon}(\bx,\omega)\right]\times
 \\
 &\left\vert\hat{\phi}(\beta,\omega,\omega_{\rm c})\Psi(\alpha,\bx,\bx_{\rm c})\right\vert^{2}
 \,{\rm d}\bx\,{\rm d}\omega
 \\
 &= C_{0}{\rm Re}\left[\hat{\varepsilon}(\bx_{\rm c},\omega_{\rm c})\right].
 \end{split}
 \end{align}
Although, we can always make $S^{\rm e}_{1}=S^{\rm e}_{2}$ by picking a suitable value for the dimensional 
coefficient $C_{0}$,
there is an important difference between the ``proper'' field $\hat{\bE}_{1}$ and the essential modes.
When we say that $\hat{\bE}_{1}$ satisfies Maxwell's equations we mean that it is a solution of a boundary value 
problem, which exists and is unique in most circumstances. The essential modes, however, are not unique.
For example, for a fixed $C_{0}$ the conservation of action holds with any $\bx_{\rm c}$ and $\omega_{\rm c}$ at the equipotential hypersurface 
(hypervolume) of ${\rm Re}\left[\hat{\varepsilon}(\bx_{\rm c},\omega_{\rm c})\right]$. 

Hence, the action-preserving transformation between the usual solutions of the Maxwell equations and the essential 
modes is nonunique in two respects. First, with a fixed value of action there are no means to decide whether 
this transformation takes place at all. Second, if such transformation has happened, then the parameters of the location 
of the spatial singular mode $\bx_{\rm c}$ and the angular frequency $\omega_{\rm c}$ of the
temporal essential mode are completely arbitrary. 

Let us see, if the physical interpretation of the essential modes can be extended any further than
the squared and integrated action-like expressions. The previous consideration suggests that in the transition 
between the states both $\bx_{\rm c}$ and $\omega_{\rm c}$ should be treated as random variables. 
Suppose that the transition is such that each essential mode can carry only a 
finite portion of action, say $a$, and, therefore, the second state corresponding to some sufficiently large numerical value of action $S^{\rm e}_{2}\gg a$
consists of a large number of essential modes rather than a single mode. 
One possibility is to assign $C_{0}=h/\varepsilon_{0}$ in (\ref{eq:EssentialAction}), describing the single-mode mode
transformation, where $h$ is the Plank constant.
Then, the amount of action ${\rm d}S^{\rm e}_{2}$ 
corresponding to the essential modes with positions $\bx_{\rm c}$ in the subvolume ${\rm d}\bx_{\rm c}$ and frequencies
$\omega_{\rm c}$ in the interval ${\rm d}\omega_{\rm c}$ is
 \begin{align}
 \label{eq:PartialSingular}
 \begin{split}
 {\rm d}S^{\rm e}_{2}=&\; C f(\bx_c){\rm d}\bx_{c}\,g(\omega_{\rm c}){\rm d}\omega_{c}
 \times
 \\
 &\lim\limits_{\alpha,\beta\rightarrow\infty}\int_{-\infty}^{\infty}\int_{\bx\in{\mathbb R}^{3}}
 {\rm Re}\left[\hat{\varepsilon}(\bx,\omega)\right]
 \times
 \\
 &\left\vert\hat{\phi}(\beta,\omega,\omega_{\rm c})\Psi(\alpha,\bx,\bx_{\rm c})\right\vert^{2}
 \,{\rm d}\bx\,{\rm d}\omega,
 \end{split}
 \end{align}
where $f(\bx_c){\rm d}\bx_{c}$ and $g(\omega_{\rm c}){\rm d}\omega_{c}$ are the relative numbers
of essential modes with parameters within the corresponding subvolume and interval. Note that $\bx_{\rm c}$
and $\omega_{\rm c}$ are still random. Hence, $f(\bx_{\rm c})$ and $g(\omega_{\rm c})$ can be interpreted
as probability densities. The complete action in the second state will look as follows:
 \begin{align}
 \label{eq:IntegralSingular}
 \begin{split}
 S^{\rm e}_{2}=&\,C \lim\limits_{\alpha,\beta\rightarrow\infty}
 \int_{-\infty}^{\infty}\int_{\bx_{\rm c}\in{\mathbb R}^{3}}f(\bx_c)\,g(\omega_{\rm c})
 \times
 \\
 &\int_{-\infty}^{\infty}\int_{\bx\in{\mathbb R}^{3}}
 {\rm Re}\left[\hat{\varepsilon}(\bx,\omega)\right]\times
 \\
 &\left\vert\hat{\phi}(\beta,\omega,\omega_{\rm c})\Psi(\alpha,\bx,\bx_{\rm c})\right\vert^{2}
 \,{\rm d}\bx\,{\rm d}\omega\,{\rm d}\bx_{c}\,{\rm d}\omega_{c}.
 \end{split}
 \end{align}
Now, equating $S^{e}_{1}=S^{e}_{2}$ again, interchanging the order 
of integration, and using the symmetry and the sifting property 
of the essential modes we obtain
 \begin{align}
 \label{eq:IntegralEquality}
 \begin{split}
 &\int_{-\infty}^{\infty}\int_{\bx\in{\mathbb R}^{3}}
 {\rm Re}\left[\hat{\varepsilon}(\bx,\omega)\right]\left\vert\hat{\bE}_{1}(\bx,\omega)\right\vert^{2}
 \,{\rm d}\bx\,{\rm d}\omega=
 \\
 &C \int_{-\infty}^{\infty}\int_{\bx_{\rm c}\in{\mathbb R}^{3}}
 {\rm Re}\left[\hat{\varepsilon}(\bx,\omega)\right]\times
 \\
 &\lim\limits_{\alpha,\beta\rightarrow\infty}\int_{-\infty}^{\infty}\int_{\bx\in{\mathbb R}^{3}}
 f(\bx_c)\,g(\omega_{\rm c})\times
 \\
 &\left\vert\hat{\phi}(\beta,\omega,\omega_{\rm c})\Psi(\alpha,\bx,\bx_{\rm c})\right\vert^{2}
 \,{\rm d}\bx_{c}\,{\rm d}\omega_{c}\,{\rm d}\bx\,{\rm d}\omega=
 \\
 &C \int_{-\infty}^{\infty}\int_{\bx_{\rm c}\in{\mathbb R}^{3}}
 {\rm Re}\left[\hat{\varepsilon}(\bx,\omega)\right]f(\bx)g(\omega)\,{\rm d}\bx\,{\rm d}\omega.
 \end{split}
 \end{align}
We also conclude that this equality holds if
 \begin{align}
 \label{eq:BornRule}
 f(\bx)g(\omega)= C^{-1}\left\vert\hat{\bE}_{1}(\bx,\omega)\right\vert^{2}.
 \end{align}
In other words, the probability density function of the random variables $\bx_{\rm c}$ and $\omega_{\rm c}$ 
is equal to the normalized squared amplitude of the electromagnetic field. This, obviously, resembles another 
postulate of the quantum theory, known as the Born rule. Although, we have to point out that 
equation (\ref{eq:IntegralEquality}) may hold for other densities as well, and more information
is required to make (\ref{eq:BornRule}) the unique choice.

Thus, Maxwell's equations admit generalized Weyl's solutions which exhibit many of the features of the 
enigmatic wavefunction collapse. Furthermore, the existence of essential modes suggests the following hierarchy of
equations in the electromagnetic field theory. There are global equations on the level of the electromagnetic action
allowing the hypothetical transitions between ``proper'' and ``essential'' states. These equations can be satisfied both by the 
usual solutions of the Maxwell equations and by the essential modes. Mathematically, such equations are simply equalities between 
quadratic functionals defined on functional spaces far broader than the common and rather restrictive complete Hilbert space. 
The more informative local (pointwise) equations involving intensities (squared amplitudes) of the field are satisfied by the ``proper'' solutions in 
the usual sense, and only statistically by the essential modes. 
The random parameters are the locations of the spatial singular modes and the angular frequency of the temporal modes. 
The natural emergence of the Born statistical rule here is truly remarkable, although, more work is required to justify it.
Finally, we have the electromagnetic field, which satisfies the Maxwell equations in the usual sense. Essential modes do not 
have any mathematical and, hence, physical meaning on that level. 

Instead of the electric field action (\ref{eq:Action})
we could use a sum of several action-like or other terms quadratic in fields, requiring the concervation of the total expression 
under the transformation of modes. Although the exact physical interpretation of the transformation may change in that case, 
the methods developed here should remain valid. Hence, we do not know yet how 
to induce or control the electromagnetic collapse in practice, and if this can be done at all. We do know the condition for the essential 
resonance, i.e.,  $\lambda^{\rm e}=-i\omega_{\rm c}\varepsilon(\bx_{\rm c},\omega_{\rm c})=0$. With ``proper'' eigenmodes the 
presence of a zero eigenvalue means that all energy is stored in the corresponding mode, which is a realizable spatial distribution of the field. 
The essential resonance, however, is not so easy to interpret. We can only discuss the relevant physical conditions, keeping in mind that $\bx_{\rm c}$
and $\omega_{\rm c}$ are random.
One straightforward case is $\omega_{\rm c}=0$, while $\varepsilon(\bx_{\rm c},\omega_{\rm c})\ne 0$, 
corresponding to the static breakdown of wave motion. Another possibility 
is $\varepsilon(\bx_{\rm c},\omega_{\rm c})=0$, while $\omega_{\rm c}\ne 0$.
This does not happen in causal media. On the other hand, with strong anomalous dispersion the real part of $\varepsilon$ may 
be zero and even negative. However, this is always accompanied by the increase in the imaginary part 
of $\varepsilon$. Which rises a question: should not we see the absorption of the electromagnetic radiation,
empirically described by the imaginary part of $\varepsilon$, as resulting from the excitation of essential modes?
The concervation of energy (Poynting theorem) may help to derive the corresponding relation.

Hence, collapse of the wave behavior is not such a crazy thing after all. Mathematically it can be viewed as the action-preserving 
transition between the states corresponding to discrete and essential spectra of operators, which does not depend on the 
loss of coherence due to environment or on the presense of nonlinearity.
Certainly, it is not clear yet how to extend these ideas on the quantum wave motion.
In any case, our experience with electromagnetics shows that the critical analysis of the mathematical conditions on the 
existence of the solution with the subsequent explicit 
derivation of the essential modes as the Weyl solutions of the first-order
Dirac's equation may be a good starting point.

\end{document}